\documentclass[]{spie}  

 \usepackage{subcaption}
\usepackage{amsmath,amsfonts,amssymb}
\usepackage{graphicx}
\usepackage[colorlinks=true, allcolors=blue]{hyperref}

\title{Effect of noise and topologies on multi-photon quantum protocols}

\author[a]{Nitin Jha}
\author[a]{Abhishek Parakh}
\author[b]{Mahadevan Subramaniam}
\affil[a]{Kennesaw State University, GA, USA}
\affil[b]{University of Nebraska Omaha, NE, USA }

\authorinfo{Further author information: (Send correspondence to Nitin Jha)\\E-mail: njha1@students.kennesaw.edu\\}

\begin{document} 
\maketitle

\begin{abstract}
Quantum-augmented networks aim to use quantum phenomena to improve detection and protection against malicious actors in a classical communication network. This may include multiplexing quantum signals into classical fiber optical channels and incorporating purely quantum links alongside classical links in the network. In such hybrid networks, quantum protocols based on single photons become a bottleneck for transmission distances and data speeds, thereby reducing entire network performance. Furthermore, many of the security assumptions of the single-photon protocols do not hold up in practice because of the impossibility of manufacturing single-photon emitters.  
 
Multi-photon quantum protocols, on the other hand, are designed to operate under practical assumptions and do not require single photon emitters. As a result, they provide higher levels of security guarantees and longer transmission distances. However, the effect of channel and device noise on multiphoton protocols in terms of security, transmission distances, and bit rates has not been investigated. In this paper, we focus on channel noise and present our observations on the effect of various types of noise on multi-photon protocols. We also investigate the effect of topologies such as ring, star, and torus on the noise characteristics of the multi-photon protocols. Our results show the possible advantages of switching to multi-photon protocols and give insights into the repeater placement and topology choice for quantum-augmented networks.  
\end{abstract}


\section{Introduction}
\label{Sec:Intro}
\label{sec:intro}  
Ever since Bennett and Brassard proposed the first protocol for quantum key distribution in $1984$, there has been rapid development of different protocols concerning the different aspects of QKD, such as efficiency, security, resource consumption, etc. \cite{bennett1992quantum, bennett2014quantum, parakh2018using, parakh2SPIE2018, parakh1SPIE2018} There have been several physical implementations of QKD protocols that span several thousand kilometers through the free space channel, several hundreds of kilometers through fiber optic cables, and several meters through underwater quantum communication channels.\cite{liao2017satellite, pirandola2020advances, xu2020secure, feng2021experimental} Experimental QKD networks such as DARPA networks, several European networks\cite{peev2009secoqc}, and Tokyo\cite{sasaki2011field} are great examples of recent developments to make QKD networks practical under current technologies. Furthermore, the four-node inter-European QKD network established during the G-20 summit held in Trieste in the year 2021 gave us an idea of a broad range QKD network as well.\cite{ribezzo2023deploying}. As with rapidly growing quantum computing devices and technologies, quantum cryptography offers security solutions and thus provides a larger-scale secure communication channel.\cite{parakhNature2022, burrUbiquity2022, burr2022evaluating}

Quantum Secure Direct Communication (QSDC) is the methodology of sending direct secret messages using principles of quantum mechanics by eliminating the middle stage involving quantum key distribution.\cite{parakh2016correcting} Various QSDC schemes have been proposed. \cite{Bostrom2002, Deng2004, Gao2008, Li2011, Long2002} Kak's three-stage protocol is the major focus of this study, and it is established to be one of the QKD protocols that is favorable to QSDC.\cite{Thapliyal2018} The two main issues restraining the development of ideal QKD setups are (1) the presence of noise in the system and (2) non-ideal photon emitters, i.e., ideal single-photon emitters are not plausible under today's technologies. The main noises that affect a practical quantum channel are amplitude damping, dephasing error, collective rotation, and bit-flip error. \cite{nielsen2010quantum} In the current Noisy Intermediate Scale Quantum (NISQ) era, we need to model such errors and develop error correction strategies to help re-envision the design of the new quantum protocols and routers. \cite{shi2023quantum} 

Kak's three-stage protocol is shown to be invariant under multiphoton implementations\cite{parakh2016correcting} In other words, the three-stage protocol does not rely on single-photon emitters for security and can tolerate large photon bursts. Therefore, the system security degrades gracefully. The theoretical effects of a few noise models on the overall efficiency of Kak's three-stage protocol have been studied earlier.\cite{Thapliyal2018} We take the study further and evaluate the performance of the three-stage protocol under several new but practical noise models. Given a noisy environment, one can argue that the tolerance of the three-stage protocol to multi-photon bursts makes it akin to a repetition error-correcting code where the same photon value is transmitted several tens of times. We will look at the strength of this error correction method under noisy environments. 

This paper studies the performance of Kak's three-stage protocol in noisy channels. Section \ref{sec:qkd} gives a brief overview of the three-stage protocol and the general effects of several noise models on the three-stage protocol. Section \ref{Sec:Simulator} defines our simulation setting and goes into depth about several noise models used for the study, and it also defines various topologies used in our simulations. Section \ref{Sec:Results} highlights the results of the study for various different noise models and in different topologies. Section \ref{Sec:Conclusion} concludes the study and highlights some of the possible research work on the related topic.

\section{QKD Protocols}
\label{sec:qkd}
Quantum key distribution provides numerous security advantages over classical key distribution methods as it relies on the laws of quantum mechanics, such as the no-cloning theorem, quantum superposition, entanglement, and observer's effect (measurement disrupts the state of the system). Many early QKD protocols, such as BB84 and B92, rely on single-photon transmissions. With a single-photon transmission, the overall security of the system can be considered high against eavesdropping attacks. This can be attributed to the fact that the information encoded on the photon cannot be cloned, and thus Eve has no way of decoding the information of this qubit (photon) without disrupting the overall state of the system and thus alerting the senders and receivers about the possible presence of an attacker in the system. However, today's technology does not allow for single-photon transmitters; thus, systems are prone to several attacks.\cite{burr2022evaluating} One such attack is known as the Photon-Number-Splitting Attack (PNS) where the eavesdropper can siphon one or more of the photons from a multi-photon burst without being noticed by the legitimate parties, as the loss of photons may be attributed to practical imperfections associated with the network.

\subsection{The Three-Stage QKD Protocol}
\label{Sec:3-stage}
The main focus of this study is the three-stage protocol\cite{Kak2006-3Stage}, which is briefly described below.
\begin{enumerate}
    \item Assume that Alice has a single-qubit quantum state $|\psi \rangle \in (\alpha|0\rangle+\beta|1\rangle)$ that she wants to securely transmit to Bob. The basis for qubit preparations is discussed between Alice and Bob beforehand and is considered global knowledge. 
    \item Alice applies a unitary operation, $U_A = R(\theta)$ to modify the state of $|\psi\rangle \to |\psi'\rangle$, where $|\psi'\rangle = U_A|\psi\rangle$. Now, Alice transmits $|\psi'\rangle$ to Bob. The unitary operation describes a rotation operation as described by eq(\ref{eq:r(theta)}).
   \begin{equation}
    R(\theta) = 
    \begin{pmatrix}
        \cos\theta & -\sin\theta \\
        \sin\theta & \cos\theta\\
    \end{pmatrix}
    \label{eq:r(theta)}
    \end{equation}
    \item Bob also applies another unitary transformation, $U_B = R(\phi)$ to transform the qubits state from $|\psi'\rangle \to |\psi''\rangle$ where $|\psi''\rangle= U_B U_A|\psi\rangle$ and transmits the new state back to Alice. One of important things to note is that $U_A$ and $U_B$ are chosen to commute, i.e., $[U_A, U_B]=0$.
    \item Due to the commuting nature of the unitary operators used in this case, Alice reverses her transformation by applying $U_A^{\dagger}$. Now, Alice transmits this updated state, i.e., $|\psi'''\rangle = U_B|\psi\rangle$, back to Bob. 
    \item Bob also reverses his unitary operation by applying $U_B^{\dagger}$. Thus, Bob recovers the original message transmitted by Alice, i.e., the initial state of qubits $|\psi\rangle$. 
\end{enumerate}
Due to the nature of the construction of the three-stage protocol, it can be easily recognised to have great potential to be used for quantum secure direct communication. There are several schemes for quantum secure direct communication, such as modified BB84 can be an ideal candidate for quantum secure direct communication if we allow Bob to have a quantum memory. \cite{Thapliyal2018} However, the three-stage protocol does not require any storage capabilities and, therefore, is practical using current technology. In the next section, we'll review the effect of some of the common-noise models over the three-stage protocol.

\subsection{Effect of Noise}
\label{Sec:Noise}
The essence of the three-stage protocol lies in the fact that the unitary matrices used by Alice and Bob commute, that is, $[U_A, U_B]=0$. However, due to the presence of noise in the system, this condition might be affected to some extent, thus reducing the overall effectiveness of the protocol. The evolution of single qubit state of the system involving noise model can be written as, \cite{preskill1998lecture, breuer2002theory, nielsen2010quantum}
\begin{equation}
    \rho = \sum_i {E^k_i\rho(E^k_i)^\dagger},
    \label{eq:state-evol}
\end{equation}
where $E^k_i$ are the respective Kraus operator for a given-respective noise models used ad $\rho$ is the density matrix representing the state of the system, and the subscript $i$ represents the different noise models, i.e., like $E_0$ defines the Kraus operator without noise application and $E_1$ represents the Kraus operator under noise application. \cite{Thapliyal2018}

\subsubsection{Commutativity of Rotation operator}
\label{Sec: Commutativity of Rotation operator}
We can write the evolution of a single-qubit quantum state for the three-stage protocol under noise assumptions as \cite{Thapliyal2018}
\begin{equation}
    \rho_k = \sum_{i,j,l}\left((U_B)^\dagger E^k_i (U_A)^\dagger E^k_j U_BE^k_l U_A \right)\rho\left( (U_B)^\dagger E^k_i (U_A)^\dagger E^k_j U_BE^k_l U_A  \right)^\dagger,
    \label{eq:state-evolution}
\end{equation}
where $k$ denotes the noise model used and $\rho = |\psi\rangle\langle\psi|$ is the initial quantum state prepared by Alice. Furthermore, $i,j,l$ denotes the independent noise model affecting the system in any of the three stages of the transmission. It can be clearly noticed that the three-stage protocol will work if and only if the Kraus operator commutes with the unitary transformations done by Alice and Bob. Considering the Kraus operator for Amplitude Damping as mentioned in Sec. \ref{Sec:Amplitude Damping Noise Model }, we can identify the cases where $[E_0, U_A]=0$.
\begin{equation}
    [E_0, U_A] = (1-\sqrt{1-p})\sin \theta \begin{pmatrix} 0 & 1 \\ 1 & 1 \end{pmatrix}
    \label{eq:AD-commutative}
\end{equation}
It's evident that eq(\ref{eq:AD-commutative}) will only vanish in the trivial cases, i.e., iff $\theta=0$ or $p=0$. Both of these cases point to the noiseless system. This points to the fact that the three-stage protocol does not work in this noisy environment in its original form. We can look at the commutativity of $E_1$ as well, 
\begin{equation}
       [E_1, U_A] = -\sqrt{p}\sin \theta \begin{pmatrix} 1 & 0 \\ 0 & -1 \end{pmatrix}
       \label{eq:AD-commutative-E1}
\end{equation}
Again from eq(\ref{eq:AD-commutative-E1}) we can see that $[E_1, U_A]=0$ \textit{iff} $p=0$ or $\theta=0$, i.e., trivial noiseless environment. \cite{Thapliyal2018}

\subsubsection{Collective Rotation Noise Effects}
\label{Sec:Collective Rotation Noise Effects}
In this section, we will explore the effect of the collective rotation (CR) noise model (as described in Sec. \ref{Sec:Random-Rotation Noise Model}) on the three-stage protocol.  Let's start with Alice sending a qubit $|X\rangle$ to Bob. Now, we consider the CR noise model has operations $U_{r1}$, $U_{r2}$, and $U_{r3}$ for the three stages involved. Thus, at the end of the three stages, the qubit received by Bob would be in state $U_{r3}U_{r2}U_{r1}|X\rangle$. We can write the expression explicitly as, \cite{parakh2016correcting}
\begin{equation}
U_{r3}U_{r2}U_{r1}|X\rangle= 
\begin{bmatrix}
e & -f \\
f & e \\
\end{bmatrix}
\begin{bmatrix}
c & -d \\
d & c \\
\end{bmatrix}
\begin{bmatrix}
a & -b \\
b & a \\
\end{bmatrix}
\begin{bmatrix}
1 \\
0 \\
\end{bmatrix}
=
\begin{bmatrix}
e(ac - db) - f(da + bc) \\
f(ac - db) + e(da + bc) \\
\end{bmatrix}
\label{eq:rotation-noise}
\end{equation}
In eq(\ref{eq:rotation-noise}), we assume that $|X\rangle = |0\rangle$ state. Assuming the random rotation angle is same for all rounds, thus final state of the qubit can be written as, 
\begin{equation}
\begin{bmatrix}
\cos\theta(\cos^2\theta-3\sin^2\theta) \\
-\sin\theta(\sin^2\theta+3\cos^2\theta)\\
\end{bmatrix}
\label{eq:random-rotation}
\end{equation}
So, the probability of error detected by Bob is given by, \cite{parakh2016correcting}
\begin{equation}
    |\sin\theta(\sin^2\theta+3\cos^2\theta)|^2
    \label{eq:noise-cr-probab}
\end{equation}

\section{Network Simulation}
\label{Sec:Simulator}
This section details our network simulator used for studying the effects of noise on the performance of the three-stage QKD protocol.

\subsection{Simulation Setting}
\label{Sec:Simulation Setting}
The simulator for this study was written in Python 3.11.4 utilizing the Qiskit library developed by IBM\cite{Qiskit}. Our study focuses on the performance of multi-photon QKD for the three-stage protocol, therefore, we start with Alice preparing her qubits in $Z$-basis, which is known by both Alice and Bob. In our simulation, we consider a bit-string of length $n$, and to model a multi-photon system, we use a total of $100$ qubits to encode each of the bits in the bit-string. This study focuses on the effects of various noises present in a practical system and thus introduces several physical noise models, such as bit-flip error, phase-flip error, attenuation error, and random-rotational errors. 

\subsection{Noise Models}
\label{Sec:Noise Model}
To make our simulator represent more practical scenarios, i.e., Noisy Intermediate Scale Quantum (NISQ)-devices, we introduced several noise models, for which we will highlight the basic mathematical definitions here. One thing to note is that all of the noise models are probabilistic in nature, i.e., the state of the qubit may or may not change due to the presence of noise in the channel. All noise models are implemented according to the probability chosen by the user, and all the errors are present in both the channel and at all the nodes, thus representing a noisy network. At all points of the communication and different stages of the three-stage protocol, there is the probability of the noise application.

\subsubsection{Amplitude Damping Noise Model}
\label{Sec:Amplitude Damping Noise Model }
One of the most predominant noise present in the system is the amplitude-damping (AD) noise, which causes the system to loose energy over time from the quantum system. This becomes relevant in the physical systems as the qubit can loose energy due to physical imperfections, i.e., to the environment and may cause the states to transition from $|0\rangle$ to $|1\rangle$. \cite{Thapliyal2018} This emission from high energy state can be modelled using Kraus Operator as, \cite{nielsen2010quantum, preskill1998lecture, thapliyal2015quasiprobability}
\begin{equation}
    E_0 = \begin{pmatrix} 1 & 0 \\ 0 & \sqrt{1-p} \end{pmatrix}, \quad E_1 = \begin{pmatrix} 0 & \sqrt{p} \\ 0 & 0 \end{pmatrix},
    \label{eq:AD_noise}
\end{equation}
where $p$ is the decoherence rate (or probability of noise affecting the system). The change in state of the system can be modelled in terms of the respective density matrices as follows, 
\begin{equation}
    \rho' = E_0 \rho E_0^\dagger + E_1 \rho E_1^\dagger,
\label{eq:AD-state}
\end{equation}
where $\rho$ is the density matrix of the quantum state before the noise application, and $\rho'$ is the change in state after the application of noise in the system. This noise model was implemented by utilizing the \textit{NoiseModel()} module from Qiskit.\cite{Qiskit}

\subsubsection{Collective-Rotation Noise Model}
\label{Sec:Random-Rotation Noise Model}
\begin{figure}[h!]
    \centering
    \includegraphics[width=0.55\textwidth]{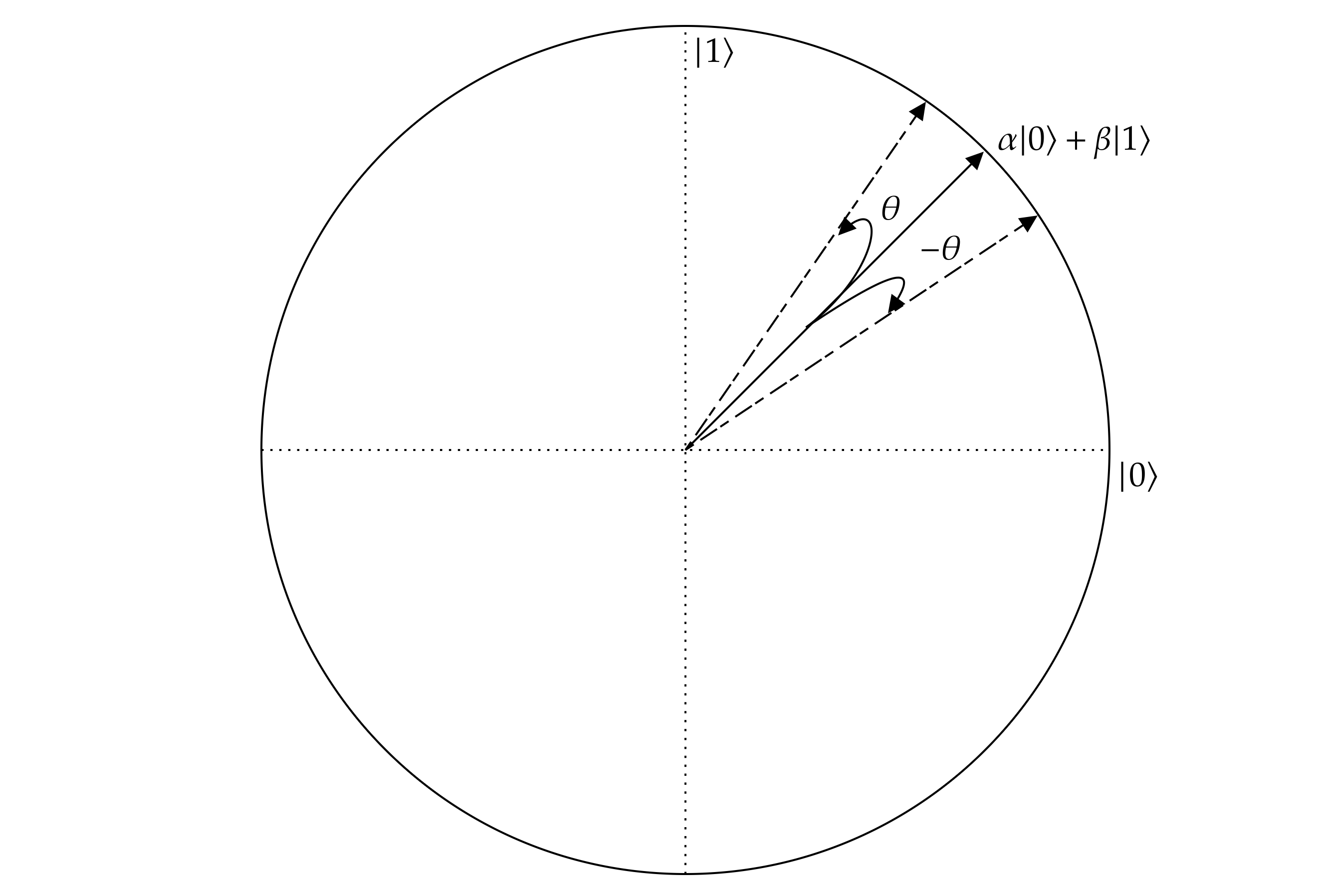}
    \caption{Depicting the random-rotational noise model on a unit circle. \cite{parakh2016correcting}}
    \label{fig:random-rotation}
\end{figure}

Apart from the amplitude damping and dephasing error, a collective random-rotational noise is the most common noise models in practical networks. Due to this noise model, a random rotational matrix ($U_r$), which is also unitary, is applied to the qubit states thus causing the following transformations, \cite{parakh2016correcting}
\begin{equation}
    |0\rangle \to U_r|0\rangle\ \& \   |1\rangle \to U_r|1\rangle
    \label{eq:rotation-noise}
\end{equation}
The states can be written in terms of the random angle of rotation, $\theta$, as following, 
\begin{equation}
|0'\rangle = \cos\theta|0\rangle + \sin\theta|1\rangle,
\label{eq:0-state-rot}
\end{equation}
\begin{equation}
    |1'\rangle = -\sin\theta|0\rangle + \cos\theta|1\rangle
    \label{eq:1-state-noise}
\end{equation}
Due to this noise mode and a random rotation applied to qubits, Bob gets the incorrect qubits as he has no way of determining this angle of random rotation. Fig(\ref{fig:random-rotation}) shows the application of a random-rotation noise model on a unit circle,\cite{parakh2016correcting}

\subsubsection{Dephasing (PD) Noise Model}
\label{Sec:phase-flip}
\begin{figure}[ht!]
    \centering
    \begin{subfigure}[b]{0.35\textwidth}
        \includegraphics[width=\textwidth]{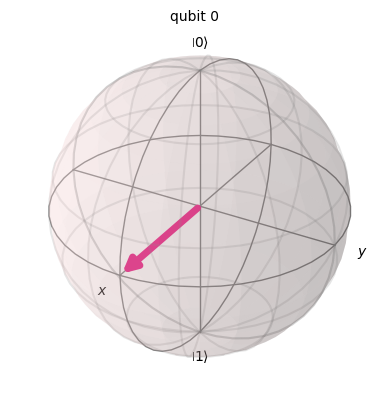}
        \caption{Qubit initiated in the state  $|\psi\rangle=(|0\rangle+|1\rangle)$, i.e., superposition of both states. Due to noise in the system, we can see the change in phase of the state of the qubit.}
    \end{subfigure}
    \hfill
    \begin{subfigure}[b]{0.35\textwidth}
        \includegraphics[width=\textwidth]{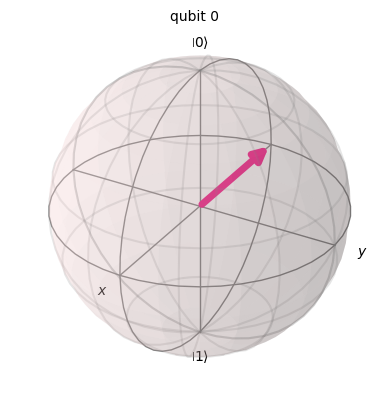}
        \caption{The phase of the qubit state being changed $|\psi\rangle=(|0\rangle-|1\rangle)$ due to the presence of dephasing noise in the system.}
        \label{1}
    \end{subfigure}
    \hfill
    \vspace{0.2cm}
    \caption{Demonstration of the effect of dephasing noise model on a qubit through Bloch sphere representation. The phase change is completely arbitrary, and this figure just serves as one of the examples of such a dephasing error in a system.}
    \label{fig:phase-flip}
\end{figure}

This noise model is analogous to the bit-flip noise model mentioned in section \ref{Sec:Bit-Flip Model}. However, phase-flip noise causes a change in the phase of the qubits, rather than being flipped from $|0\rangle$ to $|1\rangle$ or vice-versa. If we consider the probability of phase-flip being $p$, then we can write the change in density matrix as,
\begin{equation}
    \rho' = E_0 \rho E_0^\dagger + E_1 \rho E_1^\dagger,
\label{eq:phase-flip-state} 
\end{equation}
where $\rho$ is the density matrix representing the state of the qubit before the error, and $\rho'$  is the state after the error. $E_0$ and $ E_1$ are the Kraus operator for the no-error and error cases respectively and they can be described in the matrix form as given in eq(\ref{eq:E_mat_phase}).
\begin{equation}
     E_0 = \sqrt{1 - p} \begin{pmatrix} 1 & 0 \\ 0 & 1 \end{pmatrix}, \quad E_1 = \sqrt{p} \begin{pmatrix} 1 & 0 \\ 0 & -1 \end{pmatrix}
    \label{eq:E_mat_phase}
\end{equation}
To further simplify eq(\ref{eq:phase-flip-state}), we can write the equation in form of Pauli's Z-gate as, 
\begin{equation}
    \rho' = (1 - p) \rho + p Z \rho Z
\label{eq:rho_phase}
\end{equation}
where $Z$ is the Pauli-Z matrix which can be written as, 
\begin{equation}
    \begin{pmatrix} 1 & 0 \\ 0 & -1 \end{pmatrix} 
    \label{Z-gate}
\end{equation}
Fig(\ref{fig:phase-flip}) shows the effect of bit-flip noise due to which the phase of a qubit in state $|0\rangle+|1\rangle$ is changed and thus the state of the qubit becomes, for example, $|0\rangle-|1\rangle$. 

\subsubsection{Bit-Flip Noise Model}
\label{Sec:Bit-Flip Model}
\begin{figure}[h!]
    \centering
    \begin{subfigure}[b]{0.35\textwidth}
        \includegraphics[width=\textwidth]{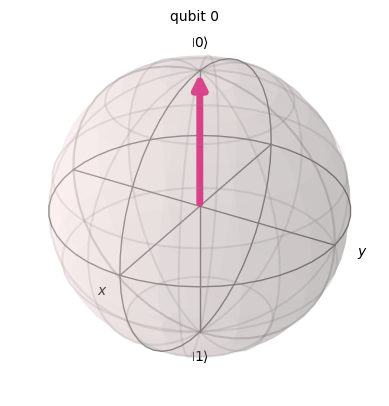}
        \caption{Qubit Initiated in $|0\rangle$ state}
        \label{fig:Direct Topology}
    \end{subfigure}
    \hfill
    \begin{subfigure}[b]{0.35\textwidth}
        \includegraphics[width=\textwidth]{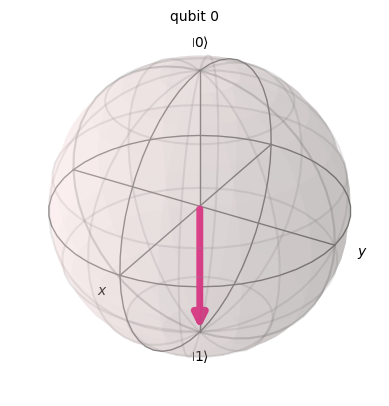}
        \caption{Qubit flipped to $|1\rangle$ due to the presence of bit-flip noise in the system.}
        \label{fig:Line Topology}
    \end{subfigure}
    \hfill
    \vspace{0.2cm}
    \caption{Demonstration of the effect of Bit-flip noise model on a qubit through Bloch-sphere representation.}
    \label{fig:bit-flip1}
\end{figure}

Similar to the classical bit-flip noise model, this noise model rotates the qubit by $180^\circ$, or \textit{flips} the state of the qubit. This noise model is equivalent to the application of Pauli's X-gate with a given probability, $p$. The state of the system post error application can be written as, 
\begin{equation}
\rho' = (1 - p) E_0 \rho E_0^\dagger + p E_1 \rho E_1^\dagger,
\label{eq:BIT-FLIP}
\end{equation}
where $\rho$ is the density matrix representing the state of the qubit before the error, and $\rho'$  is the state after the error. $E_0$ and $ E_1$ are the Kraus operator for the no-error and error cases respectively and they can be described in the matrix form as given in eq(\ref{eq:E_mat}).
\begin{equation}
     E_0 = \begin{pmatrix} 1 & 0 \\ 0 & 1 \end{pmatrix}, \quad E_1 = \begin{pmatrix} 0 & 1 \\ 1 & 0 \end{pmatrix}
\label{eq:E_mat}
\end{equation}
{Fig(\ref{fig:bit-flip1}) shows} the effect of bit-flip noise due to which a qubit initiated in state $|0\rangle$ is flipped to the state $|1\rangle$. 
\subsubsection{Bit-Phase Flip Noise Model}
\label{Sec: Bit-Phase Flip Noise Model}
The Bit-Phase flip noise model combines the two noise models described in sections \ref{Sec:Bit-Flip Model} and \ref{Sec:phase-flip}. Due to this noise, given a probability, $p$, both $X$ and $Z$ gates are applied to the qubits involved, which is considered as Pauli's $Y$-gate. The change in the state of the system can be written as, 
\begin{equation}
    \rho' = (1 - p) \rho + p Y \rho Y^\dagger,
\label{eq:rho_phase-flip}
\end{equation}
where $\rho$ is the density matrix of the qubit state before error, and $\rho'$ with the inclusion of the error and $Y$ is the Pauli's Y-gate. The Kraus operator for bit-phase flip noise model can be written as, 
\begin{equation}
    E_{Y} = \sqrt{p} \, Y,
    \label{eq:bit-phase_flip}
\end{equation}
where $p$ is the probability of the noise affecting the system and, as mentioned earlier, $Y$ is the Pauli's $Y-$gate described as follows in eq(\ref{eq:Y-gate}).
\begin{equation}
    Y = \begin{pmatrix} 0 & -i \\ i & 0 \end{pmatrix}
\label{eq:Y-gate}
\end{equation}
This noise model is implemented in our simulator by choosing a probability of applying the Pauli's Y-gate. 

\subsubsection{Attenuation Noise Model}
\label{Sec:Attenuation Noise Model}
One of the most talked about and commercially practical way to transmit protons is through optical fibers.  Optical fibers have an inherent noise which causes a loss in the total number of photons being transmitted, as the intensity of the light beams travelling through optical fibers decreases exponentially as a result of absorption and scattering losses. \cite{saleh2007attenuation} The attenuation coefficient, as denoted by $\alpha$, can be written as described in eq(\ref{eq:attenuation}). Throughout our model, the value of $\alpha$ was chosen to be $\alpha=0.15$ modeling physical systems.
\begin{equation}
    \alpha = \frac{1}{L} \left(10\log_{10}\left(\frac{1}{\tau}\right) \right),
    \label{eq:attenuation}
\end{equation}
where $\tau$ is the \textit{power transmission ratio}\cite{saleh2007attenuation}, the ratio of incident to transmitted power and $L$ is the length of the optical fiber segment. We can rearrange the terms of eq(\ref{eq:attenuation}) to get the \textit{attenuation equation}, which defines the probability of a photon being successfully transmitted over a fiber segment of length $L$ in eq(\ref{eq:atten-error}).
\begin{equation}
    P_\tau = 10^{-\alpha L/10},
    \label{eq:atten-error}
\end{equation}
where $\alpha$ is the attenuation error in $dB/km$ and $L$ is the length of the optical fiber segment between two nodes. As Qiskit\cite{Qiskit} is an ideal system simulator, i.e., no inherent loss of qubits in the system, we implement this by discarding qubits with a probability defined by eq(\ref{eq:atten-error}) for each of the bit in the given bit-string.

\subsection{Topology}
\label{Sec:Topology}
\begin{figure}[ht!]
    \centering
    \begin{subfigure}[b]{0.45\textwidth}
        \includegraphics[width=\textwidth]{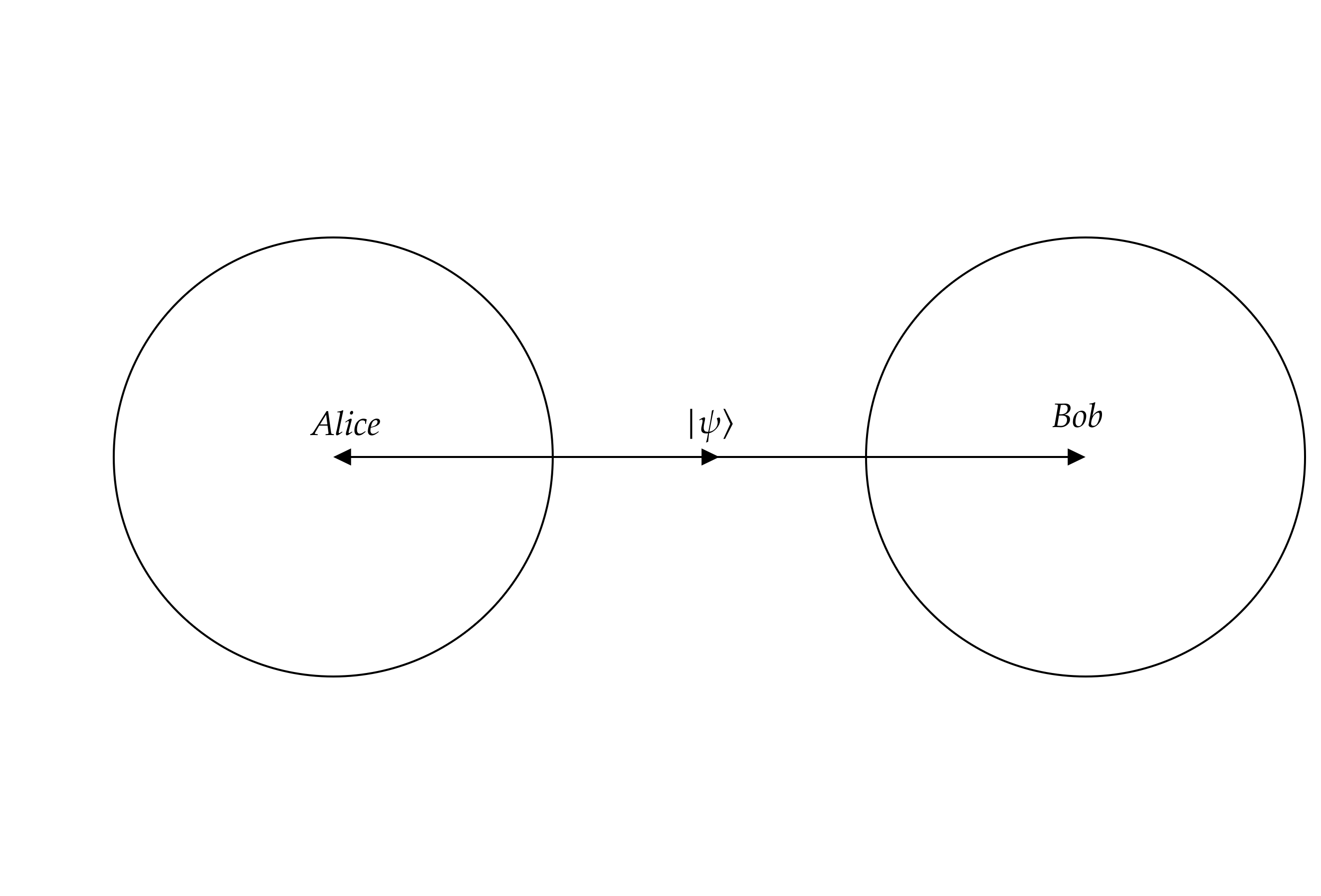}
        \caption{Schematic representation of a direct topology}
        \label{fig:direct}
    \end{subfigure}
    \hfill
     \begin{subfigure}[b]{0.45\textwidth}
        \includegraphics[width=\textwidth]{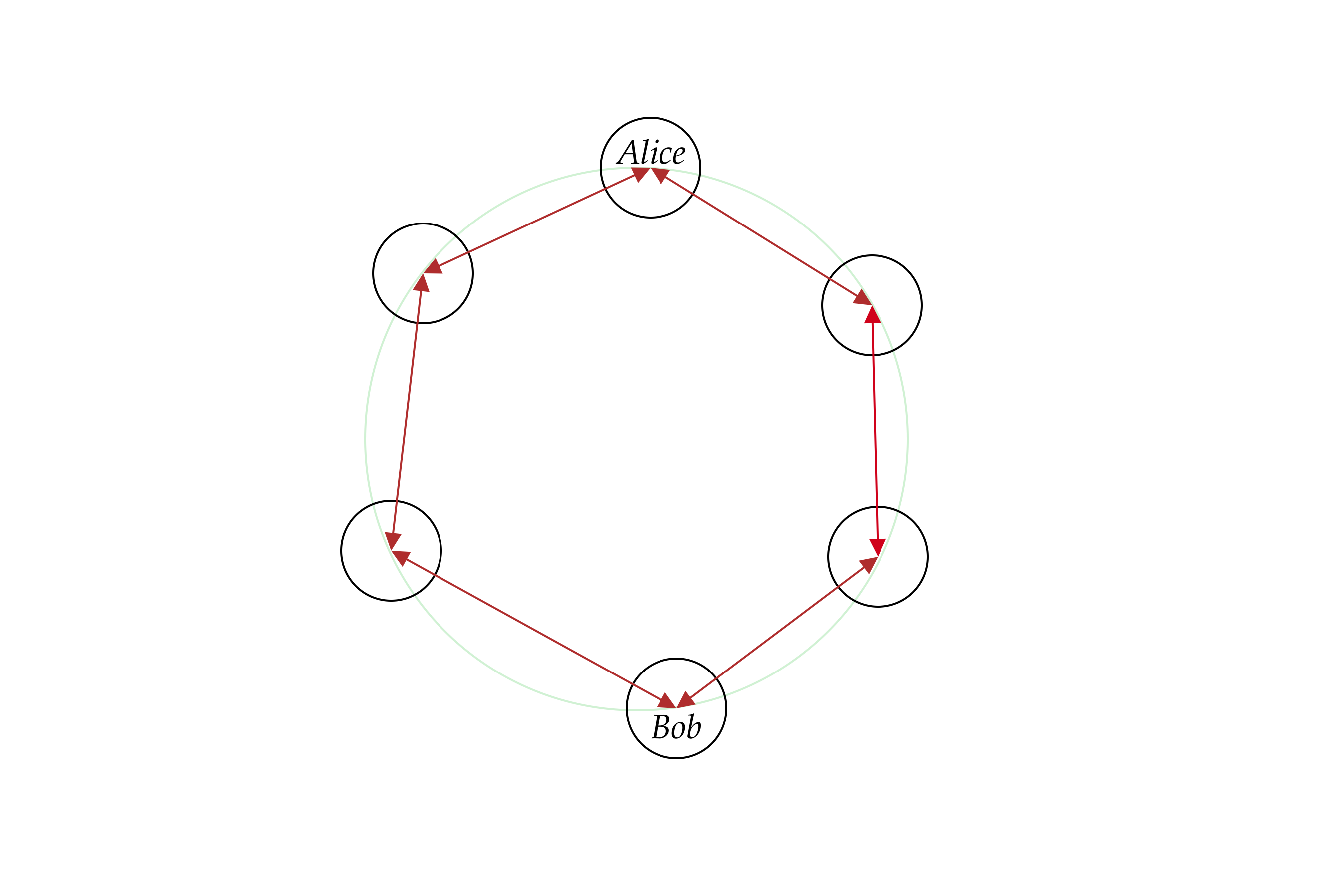}
        \caption{Schematic representation of a ring topology}
        \label{fig:ring}
    \end{subfigure}
        \hfill
     \begin{subfigure}[b]{0.45\textwidth}
        \includegraphics[width=\textwidth]{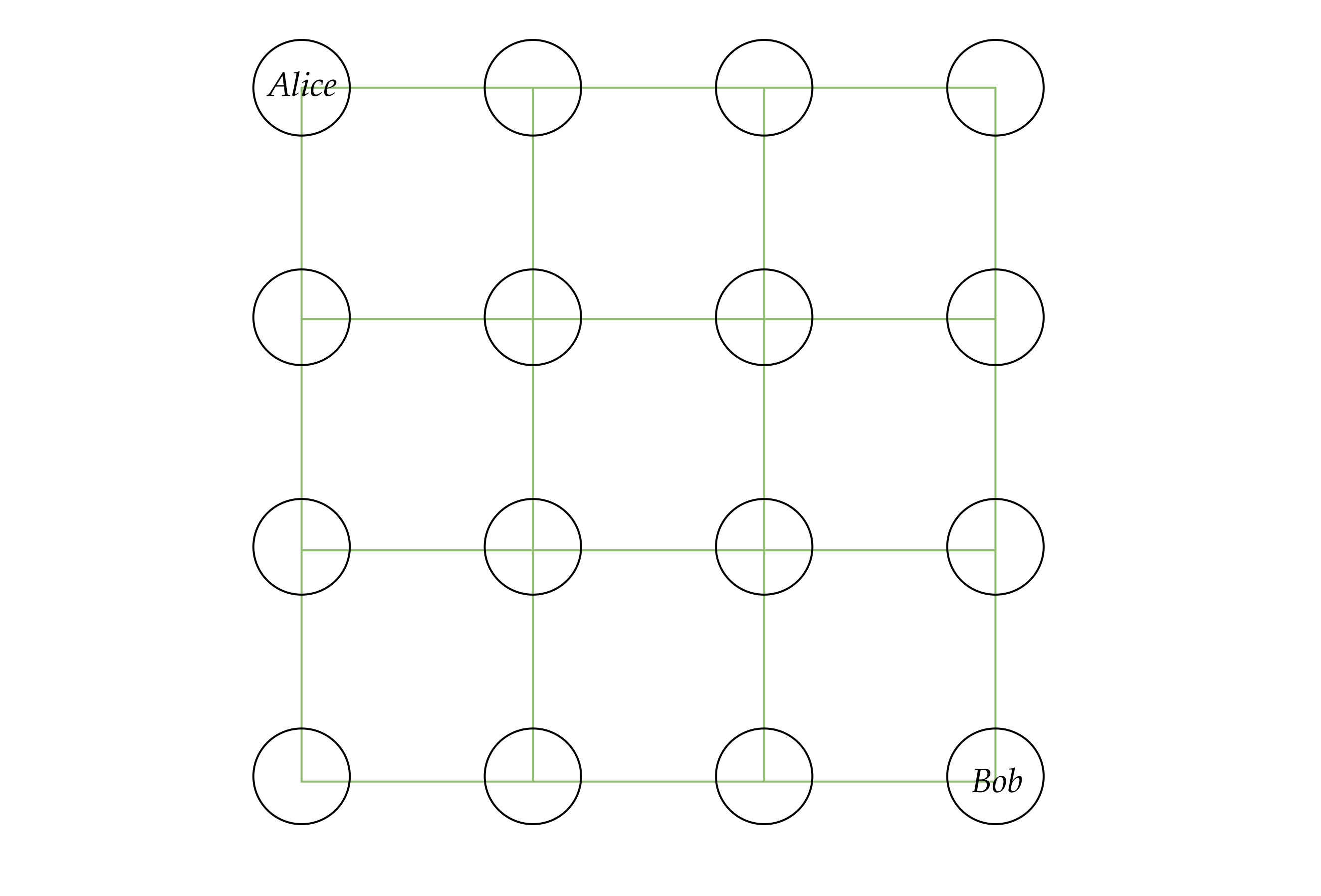}
        \caption{Schematic representation of a grid topology}
        \label{fig:grid}
    \end{subfigure}
        \hfill
     \begin{subfigure}[b]{0.45\textwidth}
        \includegraphics[width=\textwidth]{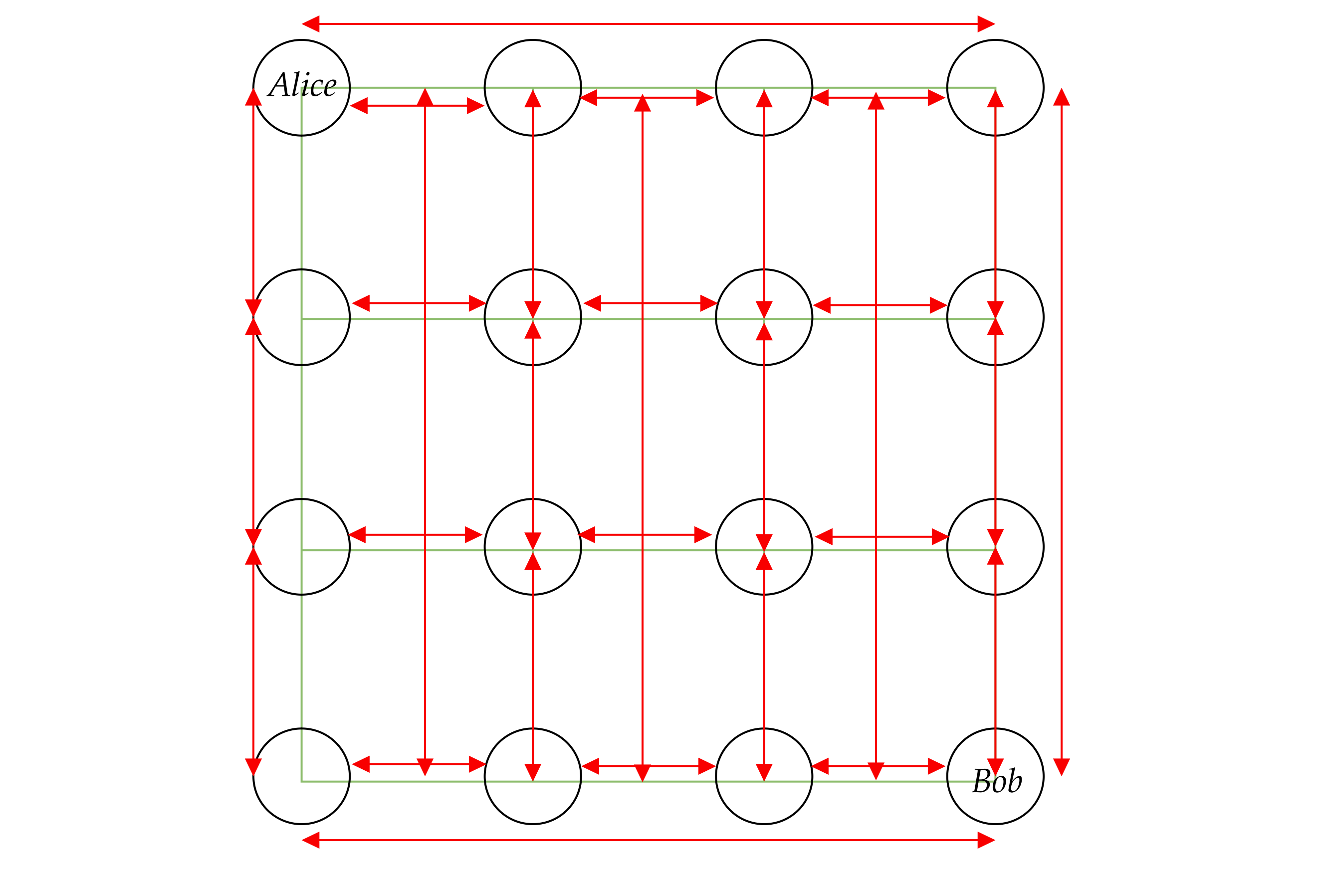}
        \caption{Schematic representation of a \(4\times 4\) torus topology}
        \label{fig:torus}
    \end{subfigure}
    \vspace{0.2cm}
    \caption{Different topologies used in our QKD simulations.}
    \label{QKD-Topologies}
\end{figure}

This paper also explores the effect of noise over different topologies while constructing the network. We implemented direct, ring, grid, and torus topology for the network parameters defining the orientation of the different nodes between Alice and Bob. As this paper deals with studying the performance of the three-state protocol under different kinds of noises in the channel network, we assume all of the nodes are trusted nodes and not prone to failures. Fig(\ref{QKD-Topologies}) represents the schematic diagrams of all of the topologies used in this study.

\subsubsection{Direct Topology}
\label{Sec:Direct Topology}
The first and our base case is the direct topology, i.e., only a direct connection between Alice and Bob with no other nodes present between them. Fig(\ref{fig:direct}) shows a schematic diagram of the direct topology used in our system.
\subsubsection{Grid Topology}
\label{Sec:Grid Topology}
The grid topology is a collection of nodes placed on a rectangular grid. Typically, there are several nodes between Alice and Bob. Alice is chosen to be the first node, and Bob is chosen to be the last node, with several intermediate nodes. As stated earlier, all nodes are trusted nodes, and there are no node failures. The distance between each node is denoted by $L$(in km), which is user-chosen. We use a breadth-first search technique (BFS) to determine the shortest distance between Alice and Bob and thus calculate the \textit{effective} distance between Alice and Bob, which contributes to the extent of attenuation noise as described in section \ref{Sec:Attenuation Noise Model}. Fig(\ref{fig:grid}) shows a schematic representation of grid topology used in our simulations. 
\subsubsection{Torus Topology}
\label{Sec:Torus Topology}
Torus topology is a grid topology with wrapping of the horizontal and vertical nodes. Torus topology is often considered a more robust topology due to the existence of multiple paths between Alice and Bob. When considering node or link failures, this comes into play and offers more robustness to the overall network performance. Fig(\ref{fig:torus}) shows a schematic representation of the torus topology used for our simulations. In fig(\ref{fig:torus}), the red lines show horizontal and vertical wrapping connections between nodes. 

\subsubsection{Ring Topology}
\label{Sec:Ring Topology}
Ring topology is a circular connection of all of the involved nodes. Due to existing of several paths for transmission, ring topology should also show more robustness, theoretically, when node and link failures are involved. However, this simulation only deals with noise models and does not take into account neither node failures nor link failures. Fig(\ref{fig:ring}) shows a schematic representation of ring topology used in our simulations. 
\section{Results}
\label{Sec:Results}
This section presents the results of the performance of the three-stage QKD protocol under various types of noise models. To calculate the success rate, we count the number of qubits at Bob's end being the same as the original bit that Alice transmitted. To decide the final bit decoded by Bob from the multi-photon qubit burst, we simply use the majority rule for that burst.

\subsection{Multi-Photon Burst Size Under Different Noise models}
\label{Sec:Multi-photon Burst size}
In this section, we change the size of multi-photon bursts, which is used to encode each of the qubits and study the performance of the three-stage protocol under a combination of different noise models such as amplitude damping with a probability of $30\%$, dephasing error with a probability of $20\%$, a collective-rotation error being applied to all qubits involved, and a phase-flip error applied with a probability of $15\%$ . In this part, we use a bit-string consisting of $96$ bits and each encoded using multiple qubits each. We change this number of qubits for each round, and study the \% of successful qubits for each bit with the presence of AD, dephasing, and CR noise models. 
\begin{figure}[h!]
    \centering
    \includegraphics[width=0.75\textwidth]{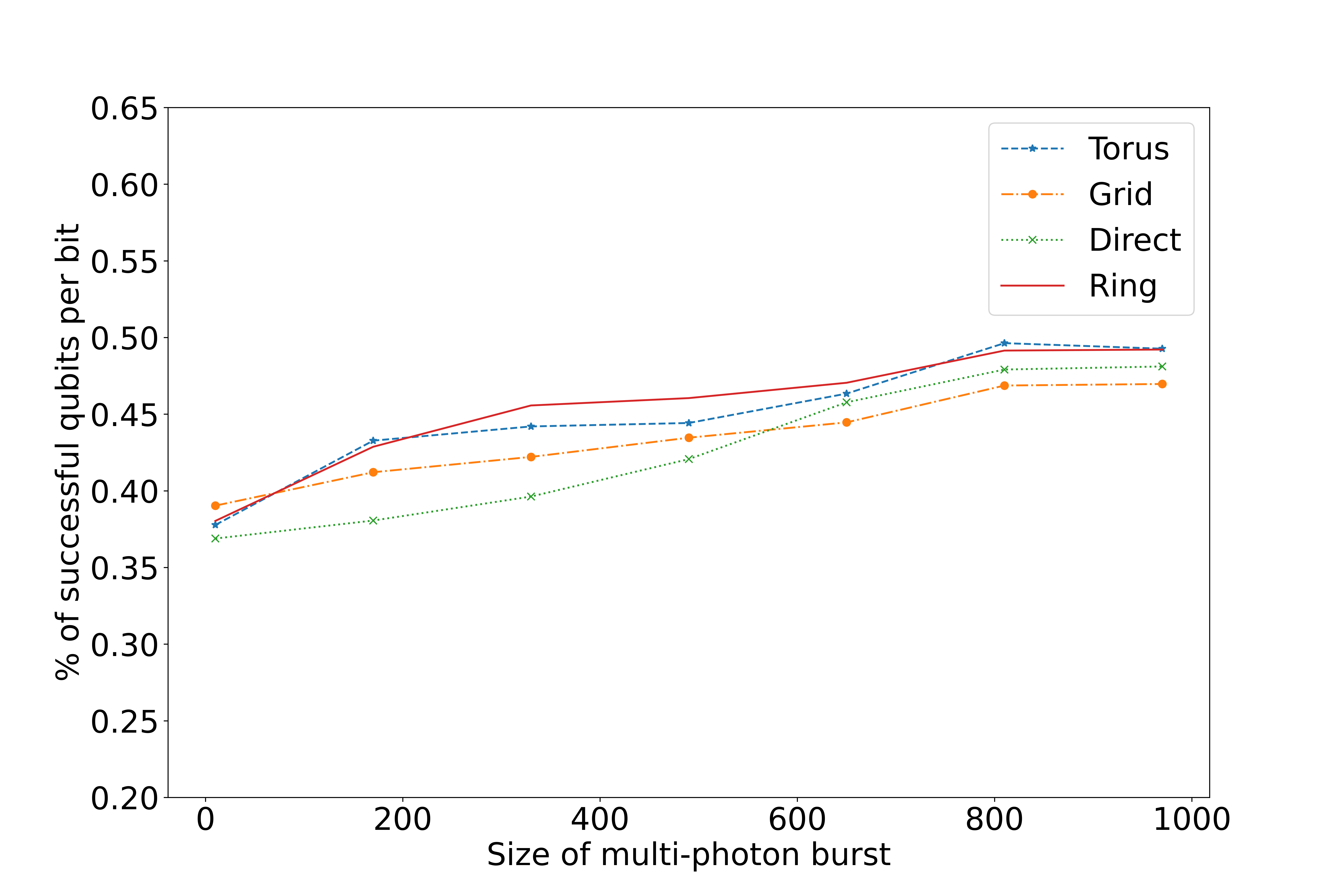}
    \caption{Relationship between multi-photon burst size and $\%$ of successful qubits per bit over different topologies.}
    \label{fig:Multi-photon}
\end{figure}

Fig(\ref{fig:Multi-photon}) shows the results between increasing multiphoton burst size and the overall count of $\%$ of successful qubits for each bit. We see that the simulated system was extremely noisy with the presence of amplitude damping noise, dephasing noise, and collective-rotation noise. However, we see that for larger burst sizes, almost for all topologies, the $\%$ of successful qubits for each bit is stabilizing around $50\%$. We also notice that for lower multiphoton burst sizes, grid topology offers better key rates, however, for higher burst sizes, torus and ring topology proves to be very robust. We also see that direct topology also shows high performance for high-burst sizes surpassing the success rates of grid topology at higher burst size which is saturating around $45\%$. From Fig(\ref{fig:Multi-photon}) we can also infer that three-stage protocol seems to be robust while utilizing the repetition correction code under multi-photon implementation for noisy environment. The performance of grid topology seems to be lesser than the other topologies present in our simulation, and this can be directly associated to the traversing of several nodes and thus an increase in error-amplification through the network run, and even after average over several runs, the performance is still weaker than the others. However, we can also see that the \% of successful qubits for all of the topologies seems to be saturating at higher multi-photon burst size ranges, highest being that of Ring and Torus around $50 \%$.  
\subsection{Different Noise Models}
\label{Res:Noise Models}
In this section, we compare the results based upon applying only one noise model at a time. 
\subsubsection{Amplitude Damping Noise Model}
In this section, we only apply amplitude damping noise model with a probability of $30$\% and study the overall efficiency of three-stage protocol by looking at \% of successful qubits for each bit. 
\begin{figure}[h!]
    \centering
    \includegraphics[width=0.65\textwidth]{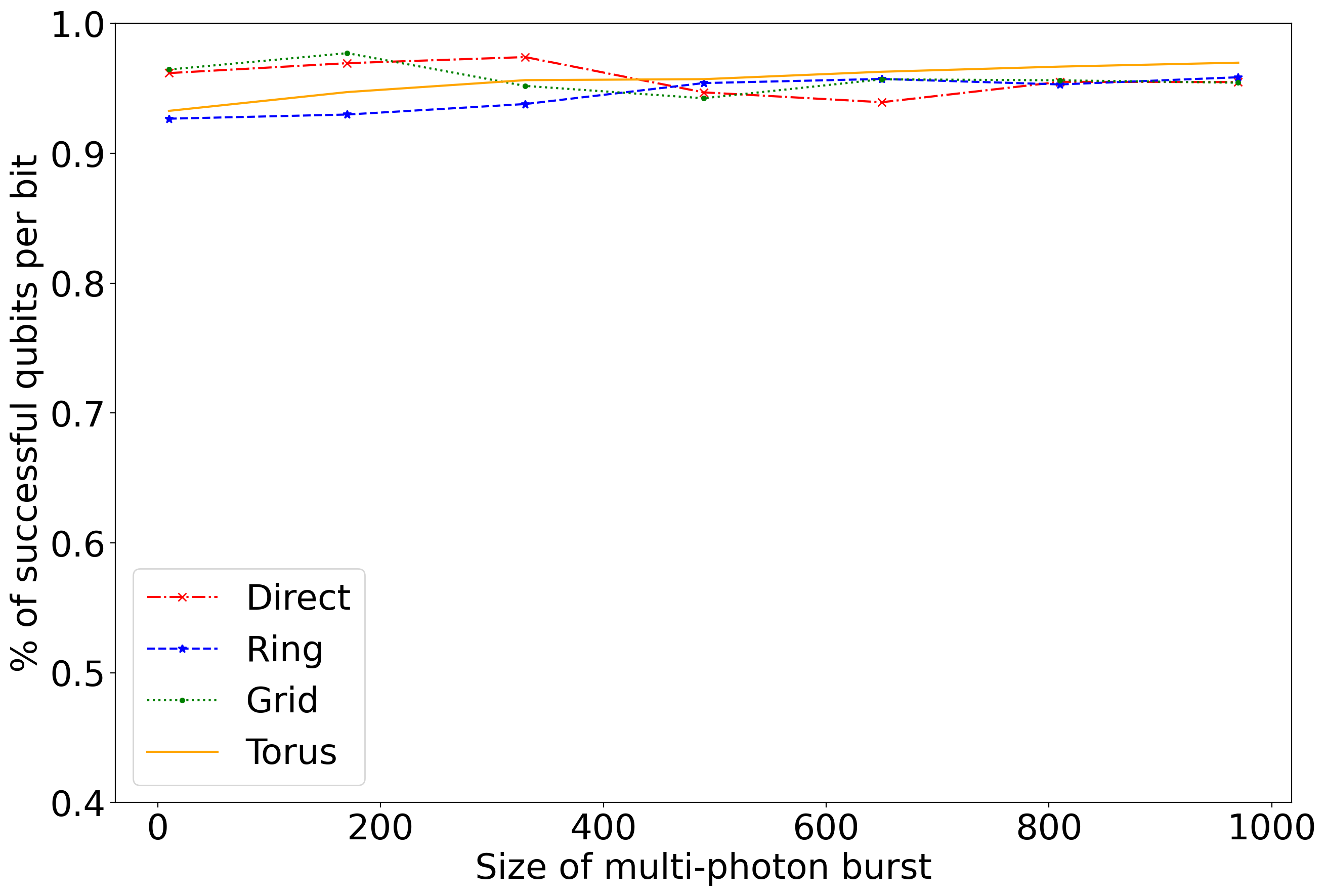}
    \caption{\% of successful qubits with different sizes of multi-photon bursts sizes over different topologies in the presence of amplitude-damping noise model applied with a probability of $20\%$.}
    \label{fig:ad-noise}
\end{figure}

From Fig(\ref{fig:ad-noise}) we can see that the $\%$ of successful qubits seems to be very consistent over all topologies around $95\%$. We can see that while the success rate is fluctuating, taking average over several network runs has given us a better estimate of the performance under probabilistic noise models application. We notice that the success rates of the qubits are between $90\%$ and $100\%$ for all of the topologies used. This gives us a positive indication about using multiphoton implementation towards correcting the possible effects of noise model.

\subsubsection{Bit Flip Noise Model}
Bit-flip error is one of the \textit{drastic} errors that can be present in the channel, as the effects of it are extreme compared to the other noise models. Here, we applied bit-flip noise models to qubits with a probability of $30\%$ again and analyze the efficiency of three-stage protocol by again looking at the $\%$ of successful qubits per bit in the bit string.

\begin{figure}[h!]
    \centering
    \includegraphics[width=0.6\textwidth]{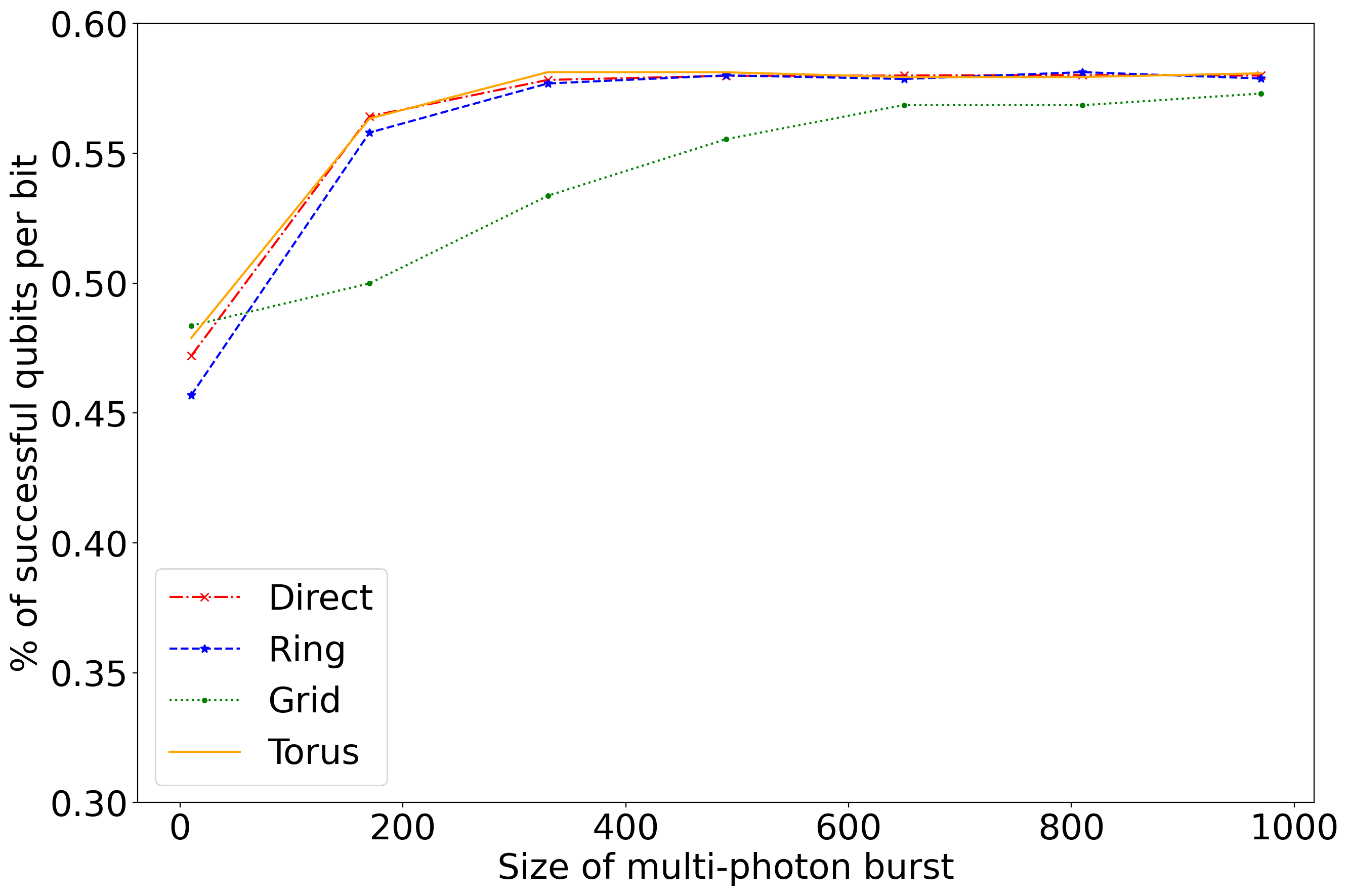}
    \caption{\% of successful qubits over different topologies for different multi-photon burst sizes under the presence of bit-flip error model. }
    \label{fig:bit-flip}
\end{figure}

From fig(\ref{fig:bit-flip}) a similar behavior to earlier presented in fig(\ref{fig:Multi-photon}), but one difference to be noted is that there were several other noise models associated in that simulation. Thus, it supports that the effects of bit-flip noise model is extreme, however by using multi-photon implementation of Kak's three-stage protocol we see that the repetition method, indeed, counters the problem cause by bit-flip noise model too. We notice that almost all of the topologies have over $50\%$ (around $58\%$) successful qubits for larger multi-photon burst size, thus ensuring that Bob will decode the messages correctly.

\subsubsection{Attenuation Noise Model}
This noise model arises due to the use of fiber-optics cable for transmission and as our simulations assumes using optical fibers for intermediate connections, we have to analyze the affects of this error on the overall performance of the system. As described in Sec.\ref{Sec:Attenuation Noise Model}, due to this error, there's a loos of qubits while transmitting from one nodes to other as an exponential function dependent on the distance between the nodes. We increase the distance between the nodes, and study the results due to it. 

\begin{figure}[h!]
    \centering
    \begin{subfigure}[b]{0.45\textwidth}
        \includegraphics[width=\textwidth]{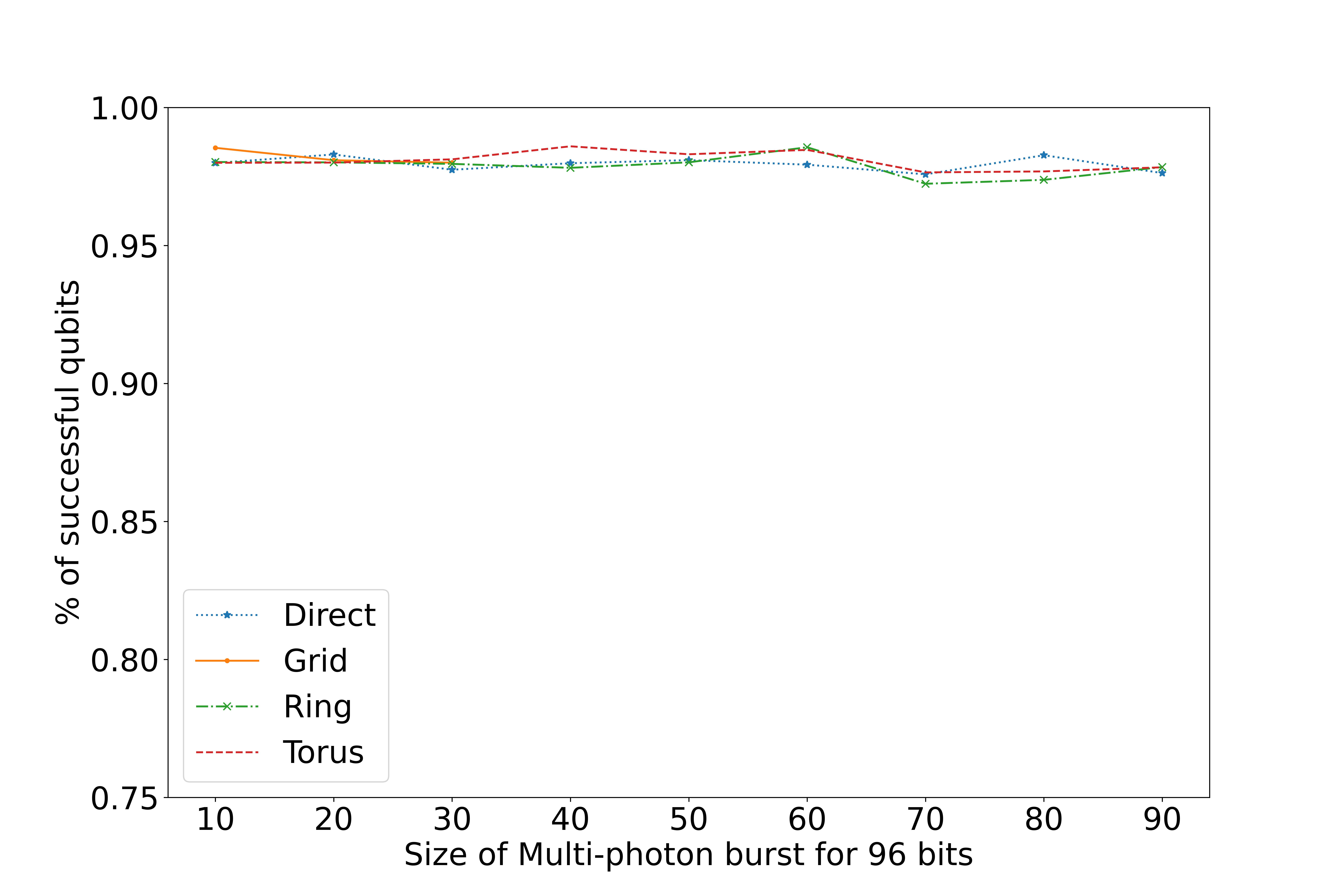}
        \caption{\% of successful qubits over different distances in kms, loss caused by attenuation error.}
        \label{fig:attenuation1}
    \end{subfigure}
    \begin{subfigure}[b]{0.45\textwidth}
        \includegraphics[width=\textwidth]{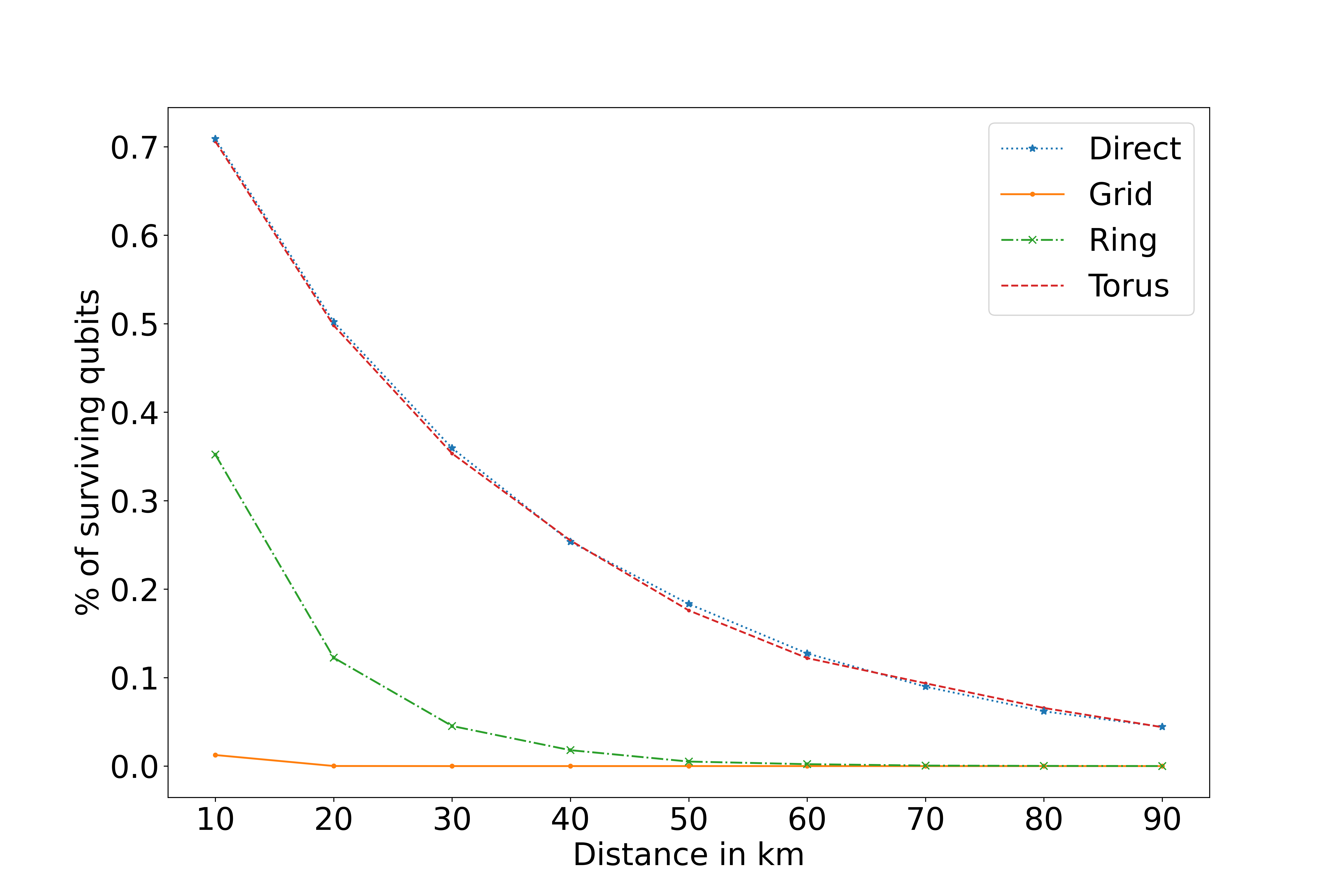}
        \caption{\% of surviving qubits being decoded by Bob, loss caused by attenuation error.}
        \label{fig:attenuation2}
    \end{subfigure}
    \caption{Results for attenuation error showing larger-loss of qubits over larger distances. }
    \label{fig:attenuation}
\end{figure}

From Fig(\ref{fig:attenuation}), we can see that the performance of all topologies were good under attenuation error. We can notice that grid topology fails significantly over larger distances between each node as the qubit has to travel a larger distance thus accumulating more optical loss.  

\subsubsection{Phase Change and Flip Noise Model}
In this section, we introduce dephasing and flip noise model to the system with a probability of $30\%$ and study the overall efficiency of three-stage protocol by looking at \% of successful qubits for each bit. In this model, as discussed in Sec.\ref{Sec:phase-flip}, there is a probability of both the dephasing and flip error, i.e., application of Pauli's Y-gate. 
\begin{figure}[h!]
    \centering
    \includegraphics[width=0.6\textwidth]{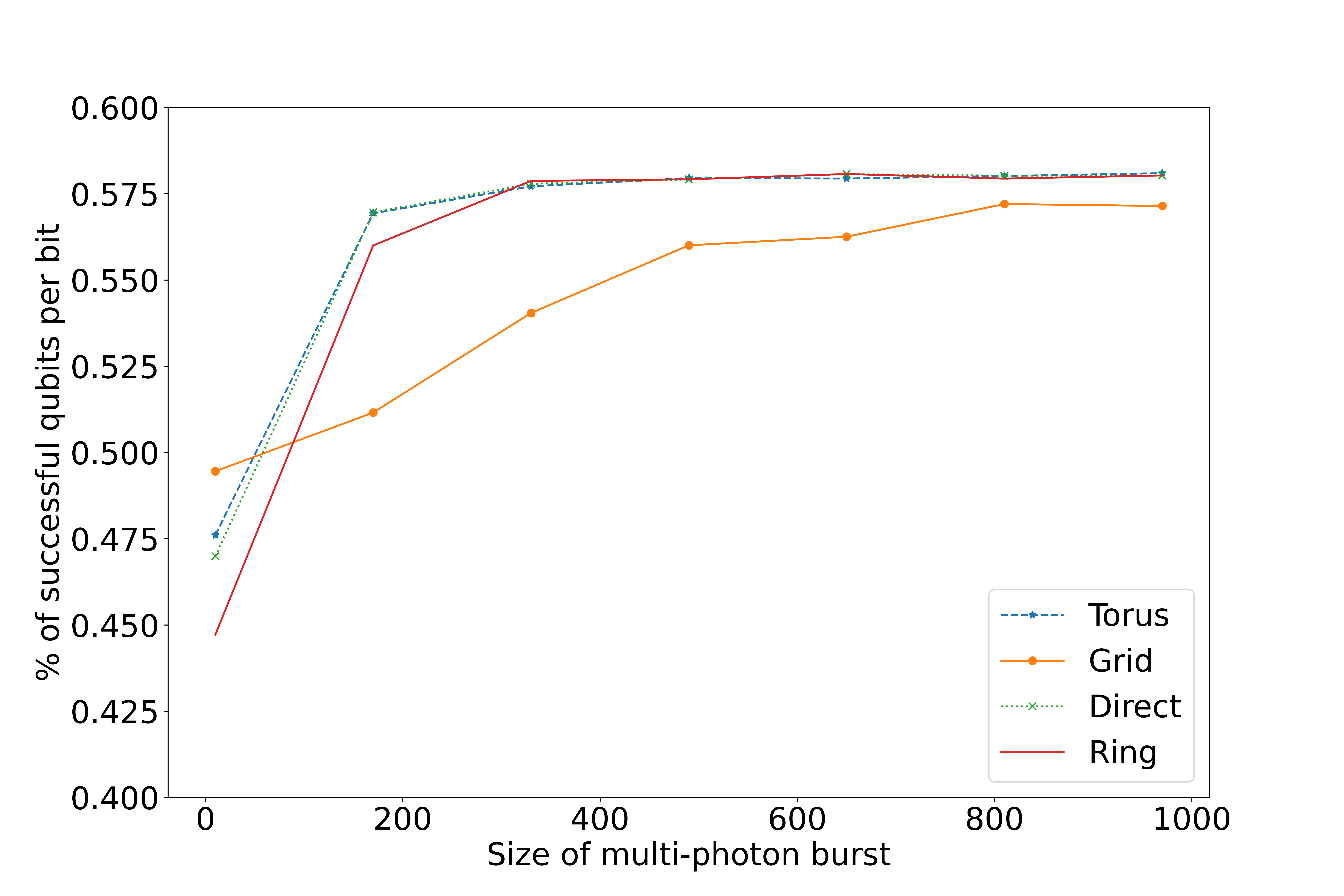}
    \caption{\% of successful qubits over different topologies for different multi-photon burst sizes under the presence of dephase and flip error model, which essentially is the application of Pauli's Y-Gate. }
    \label{fig:phasechange-flip}
\end{figure}

Fig(\ref{fig:phasechange-flip}) shows a similar nature to the graph graph in fig(\ref{fig:bit-flip}). The grid topology seems to show a gradual increase in the success rate of the qubits, and shows a high potential for higher multi-photon burst sizes. We see that all of the other topologies gains advantage over grid topology under higher multi-photon burst size. We again notice that the success rate for qubits tends to stabilize around $58\%$ for torus, ring, and direct topology for higher order burst sizes used.

\section{Conclusion}
\label{Sec:Conclusion}
QKD protocols have been of great interest in the recent past due to the rapid development of quantum computing devices. However, as some of the development has hit some obstacles due to the presence of several noises in the system, it is important to study the effects of several noise models on the popular quantum protocols. Few of the major noise models have amplitude-damping, bit-flip, dephasing, and phase-flipping error. In this study, we simulated Kak's three-stage protocol under noisy environment over various different topologies. We use multi-photon implementation, i.e., we encode each bit in the bit-string using multiple-qubits for each of the bits. We implement amplitude-damping, dephasing, attenuation-error, bit-flip, and collective-rotation error. The results of the study show the drastic effect of noise on the performance of the three-stage protocol, however due to multi-photon implementation we see a positive result favoring the use of three-stage protocol in next-generation practical networks. Direct and torus topology shows high success rates over higher multiphoton burst size; however, direct topology is very impractical while designing practical quantum-networks. Furthermore, we also examined the performance of three-stage protocol under individual noise models too. We found out that the performance of the protocol under bit-flip and phase-change/flip error resembles the performance of the protocol under the presence of majority of all of the noise models examined in Sec.\ref{Sec:Noise Model}. 

This study makes it clear that the performance of various QKD protocols needs to be studied under noisy environment. Furthermore, comparison of performance of several QKD protocols such as Coherent One-Way (COW) protocol can be of interest as an extension to this study. To summarize, this study looked at the performance of Kak's three-stage quantum protocol for multi-photon implementation and under various noise models. The study showed positive results about possible usage of three-stage QKD protocol under multi-photon implementation in practical networks utilizing the non-ideal emitters available in current age.

\bibliography{report} 
\bibliographystyle{spiebib} 

\end{document}